\documentstyle[prd, aps, preprint, epsf]{revtex} 
\begin{document} 
\draft
\title{Quantum Creation of Black Hole by Tunneling in Scalar Field 
Collapse} 
\author{Dongsu 
Bak\footnote{Electronic address: dsbak@mach.uos.ac.kr}$^{a}$, 
Sang-Pyo Kim$^{b}$, Sung-Ku Kim$^{c}$, 
Kwang-Sup Soh$^{d}$, and Jae-Hyung Yee$^{e}$} 
\address{a Department of Physics, University of Seoul, Seoul 
130-743 Korea\\ 
b Department of Physics, Kunsan National University, Kunsan 
573-701 Korea\\ 
c Department of Physics, Ewha Women's University, Seoul 120-750 
Korea\\ 
d Department of Physics Education, Seoul National University, 
Seoul 151-742 Korea\\ 
e Department of Physics, Yonsei University, Seoul 120-749 Korea}
\tightenlines 
\maketitle 
\begin{abstract} 
{ Continuously self-similar solution of  spherically symmetric 
gravitational collapse of a scalar field is studied to investigate 
quantum mechanical black hole formation by tunneling in the 
subcritical case, where, classically, the collapse does not produce 
a black hole.
} 
%PACS numbers : 04.70.Bw, 05.70.Jk 
\end{abstract}
\section{INTRODUCTION} 
Recently gravitational collapse of scalar fields have attracted 
much attention after the surprising discovery  of 
critical phenomena by 
Choptuik\cite{choptuik}. The spherically symmetric collapse of a scalar 
field has been extensively investigated both 
analytically\cite{christodoulou} and 
numerically\cite{choptuik,goldwirth}. 
The critical solution often displays 
self-similarity such that there exists a vector field, $\xi$, 
whose Lie derivative of the spacetime metric satisfies 
$L_{\xi}g=2g$. The first known analytic solution of this kind was 
found by Roberts\cite{robert} in the context of counter-examples to cosmic 
censorship, and later discussed by Brady\cite{brady} in connection with 
critical behavior. Frolov\cite{frolov} extended Roberts' work in general 
spacetime dimensions. 
 
In the work of Roberts-Brady-Frolov\cite{robert,brady,frolov} 
they considered 
subcritical, critical, and supercritical solutions depending on a 
parameter characterizing the solutions. In the supercritical 
solution a black hole is formed by the collapse, and in the 
critical case the spacetime is asymptotically flat but contains a 
null, scalar-curvature singularity. In the subcritical 
case, classically, the scalar field collapses, 
interacts and disperses leaving 
behind flat spacetime without forming a black hole. 
 
In this paper we reconsider the subcritical case to investigate the 
quantum mechanical black hole formation by tunneling. Although the 
flux of the scalar field is not intense enough to classically 
form a black 
hole, it is possible to tunnel through the effective potential 
barrier and to continue collapsing to form a black hole. In the 
classically forbidden region we shall solve the field equations 
in Euclidean spacetime, whereas Roberts, Brady, and Frolov 
considered only spacetime with Lorentzian signature. We shall show 
that the black hole does form in the subcritical case through the 
tunneling mechanism. The mass of the black hole turns out to 
be infinity due to the self-similar nature of the solution. But one 
may view the solution as a near-critical solution with a 
finite mass in the spacetime region of a scale  much below 
the relevant large correlation length. 
 
In the section II, we present the field equations and effective one 
particle formalism, and in the section III we consider solutions 
in the Lorentzian and Euclidean regions of spacetime. There are two 
kinds of Lorentzian spacetime region : one before tunneling which 
is the one studied by Roberts, Brady, and Frolov. The other is the 
region after tunneling. In the Euclidean region an instanton type 
solution is obtained. In the section IV we join continuously the 
above three regions to treat black hole formation by quantum 
tunneling of the subcritical collapse of a scalar field. In the 
last section, we discuss similar problems in general 
dimensions. 

\section{EQUATIONS OF SPHERICALLY SYMMETRIC SELF-SIMILAR FIELDS} 
We study the minimally coupled scalar 
field in the four dimensional spacetime whose action is 
\begin{equation} 
 S = \frac{1}{16\pi} \int_{M}d^{4}x 
                \sqrt{-g}~[R-2(\bigtriangledown \phi)^{2}] +
{1\over 8\pi} \int_{\partial M}  d^3x K \sqrt{-h} 
\label{eq01} 
\end{equation} 
where $K$ is the trace of second fundamental form of the boundary. 
Since we are only interested in spherically symmetric solutions, 
we can reduce the action as 
\begin{equation} 
S_{sph} = \frac{1}{4} \int \ d^{2}x 
               \sqrt{-\gamma} ~r^{2} [R(\gamma)+\frac{2}{r^{2}} 
               \{ (\bigtriangledown r)^{2} +1 \}-2(\bigtriangledown 
               \phi)^{2} ], 
\label{eq02} 
\end{equation} 
where $\gamma_{ab}$ is the metric in the remaining two-dimensional 
manifold. It is most easily handled in terms of null coordinates 
such that the metric is expressed as 
\begin{equation} 
ds^{2} = -2 e^{2 \sigma} du dv + r^{2} d \Omega^{2}, 
\label{eq03} 
\end{equation} 
where $\sigma$ and $r$ are functions of both $u$ and $v$, and $d\Omega^{2}$ 
is the line element on the unit sphere. The Einstein-scalar field 
equations read 
\addtocounter{equation}{1} 
$$ 
(r^{2})_{,uv} = -e^{2\sigma} 
\label{eq04a} 
\eqno{(4.a)}
$$ 
$$2\sigma_{,uv}-\frac{2r_{,u}r_{,v}}{r^{2}}=
\frac{e^{2\sigma}}{r^{2}}-2\phi_{,u}\phi_{,v} \eqno{(4.b)} 
$$ 
$$ 
r_{,vv}-2\sigma_{,v}r_{,v}=-r(\phi_{,v})^{2} \eqno{(4.c)} 
$$ 
$$ 
r_{,uu}-2\sigma_{,u}r_{,u}=-r(\phi_{,u})^{2} \eqno{(4.d)} 
$$ 
$$ 
2\phi_{,uv}+(\ln r^{2})_{,v}\phi_{,u}+(\ln r^{2})_{,u}\phi_{,v}=0, 
\eqno{(4.e)} 
\label{eq04e}
$$ 
where the last equation is the wave equation for the scalar field 
$\phi$, and a comma (,) denotes a partial derivative. 
 
In order to find a continuously self-similar solution, 
we impose 
the conditions as 
\begin{equation} 
\sigma(u,v)=\sigma(z),~~r=-u\rho(z),~~\phi=\phi(z),
\label{eq05}  
\end{equation} 
where z is the scale-invariant variable defined as 
\begin{equation} 
z=-\frac{v}{u}\,.
\label{eq06} 
\end{equation} 
The influx of the scalar field is turned on at the advanced time 
$v=0$, so that the spacetime is flat to the past of this surface, 
and the initial conditions are specified by continuity. The region 
of interest is the sector $u<0$, $v>0$, where we choose signs such 
that $z>0$, $\rho>0$. With this choice of the self-similar metric 
and field, the equations (4.a)-(4.e) become 
\addtocounter{equation}{1} 
$$ 
\rho^{''}\rho z+(\rho^{'})^{2}z-\rho^{'}\rho = - 
\frac{1}{2}e^{2\sigma}, \eqno{(7.a)} 
$$ 
$$ 
\sigma^{''}z+\sigma^{'} + \frac{\rho^{''}z}{\rho} = 
-z(\phi^{'})^{2}, \eqno{(7.b)} 
$$ 
$$ 
\rho^{''} - 2\rho^{'}\sigma^{'} = -\rho(\phi^{'})^{2}, 
\eqno{(7.c)} 
$$ 
$$ 
\rho^{''}z+2\sigma^{'}\rho-2\sigma^{'}\rho^{'}z=-\rho 
z(\phi^{'})^{2}, \eqno{(7.d)} 
$$ 
$$ 
\phi^{''}\rho+2\phi^{'}\rho^{'}=0. \eqno{(7.e)} 
$$ 
Prime denotes the derivative with respect to $z$. 
\begin{figure}
\epsfxsize=2.7in
\centerline{
\epsffile{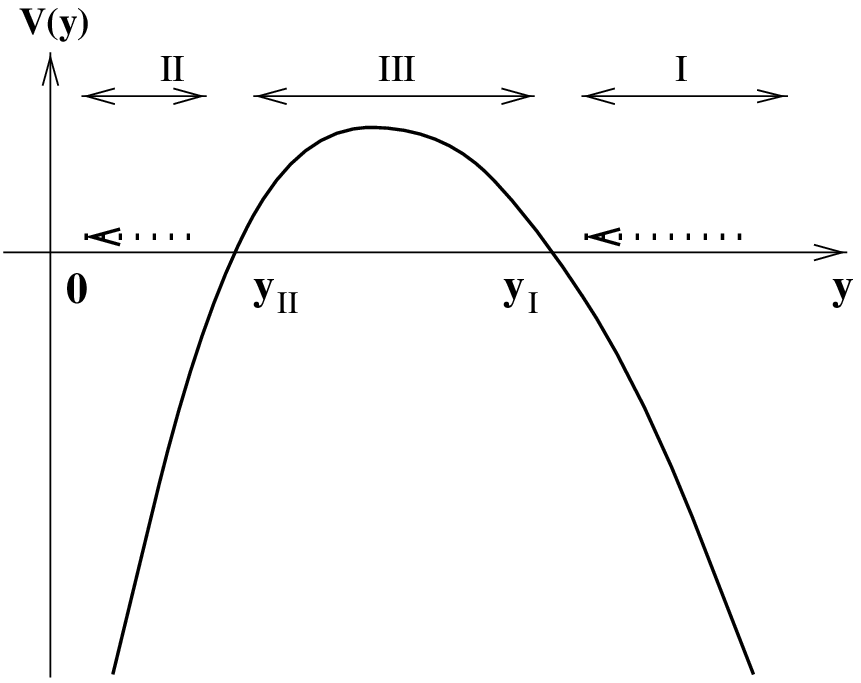}
}
\vspace{.1in}
{\small Figure~1:~The height of the potential $V(y)$ is controlled by the value
of $c_0$. The 
regions I, II and III are specified.}
\end{figure}

With new 
variables $x$ and $y$, given by 
\begin{equation} 
x=\frac{1}{2} \ln z,~~\rho=y(x)\sqrt{z}\,,
\label{eq08}  
\end{equation} 
the above equations reduce to, with the metric factor $\sigma=0$, 
\begin{eqnarray} 
&&\dot{y}^{2}=y^{2}-2+\frac{c_{0}^{2}}{y^{2}}\,,
\label{eq09}\\  
&&\dot{\phi}=\frac{c_{0}}{y^2}\,,
\label{eq10}  
\end{eqnarray} 
where $c_{0}$ is an integration constant, and dot denote 
derivative with respect to the new variable $x$. 

The equation (\ref{eq09}) for $y$ formally describes motion of 
a particle with zero energy in 
a potential 
\begin{equation} 
V(y)=2-y^{2}-\frac{c^2_{0}}{y^2}\,.
\label{eq11}  
\end{equation} 
We are interested in the production of a black hole by quantum 
tunneling in the subcrictical case, $0<c_{0}<1$, where the 
potential, only allows the classical motion of the particle starting 
from $y=\infty$ to a minimum value $y_{I}$. We call 
this the Lorentzian region I, where Roberts, Brady, 
and Frolov\cite{robert,brady,frolov} provide explicit 
solutions in this region. There is another region, we 
call the Lorentzian region II, where classical motion 
is allowed, for $0<y<y_{II}$. For quantum tunneling 
we need to 
consider the classically forbidden Euclidean region, 
$y_{II}<y<y_{I}$ as 
shown in the Fig. 1.

\section{SOLUTION IN THE LORENTZIAN AND THE EUCLIDEAN SECTORS} 

 (A) SOLUTION IN THE LORENTZIAN REGION I $(y_{I}\leq y<\infty)$ 
\vskip 0.2cm 
In this region where $y$ is restricted as $y_{I}\leq y<\infty$, 
the solutions to 
the equations (\ref{eq09}) and (\ref{eq10}) are 
\begin{eqnarray} 
&&y^{2} = 1+\sqrt{1-c_{0}^2}\cosh2x\,,
\label{eq12} \\ 
&&\phi = \frac{1}{2c_0} 
\ln\Biggl(\frac{e^{2x}+\Gamma}{e^{2x}+\Gamma^{-1}}\Biggr)-
\frac{1}{c_0}\ln\Gamma\,, 
\label{eq13} 
\end{eqnarray} 
where $0<c_{0}<1$, and $\Gamma = \sqrt{(1-c_{0})/(1+c_{0})}$. 
We note that the 
minimum value of $y$ is given by 
\begin{equation} 
y^{2}_{I}=1+\sqrt{1-c_{0}^2} ,~~x=0\,,
\label{eq14}  
\end{equation} 
but there does not occur an apparent horizon as shown by 
Brady\cite{brady} and Frolov\cite{frolov}.
\begin{figure}
\epsfxsize=2.4in
\centerline{
\epsffile{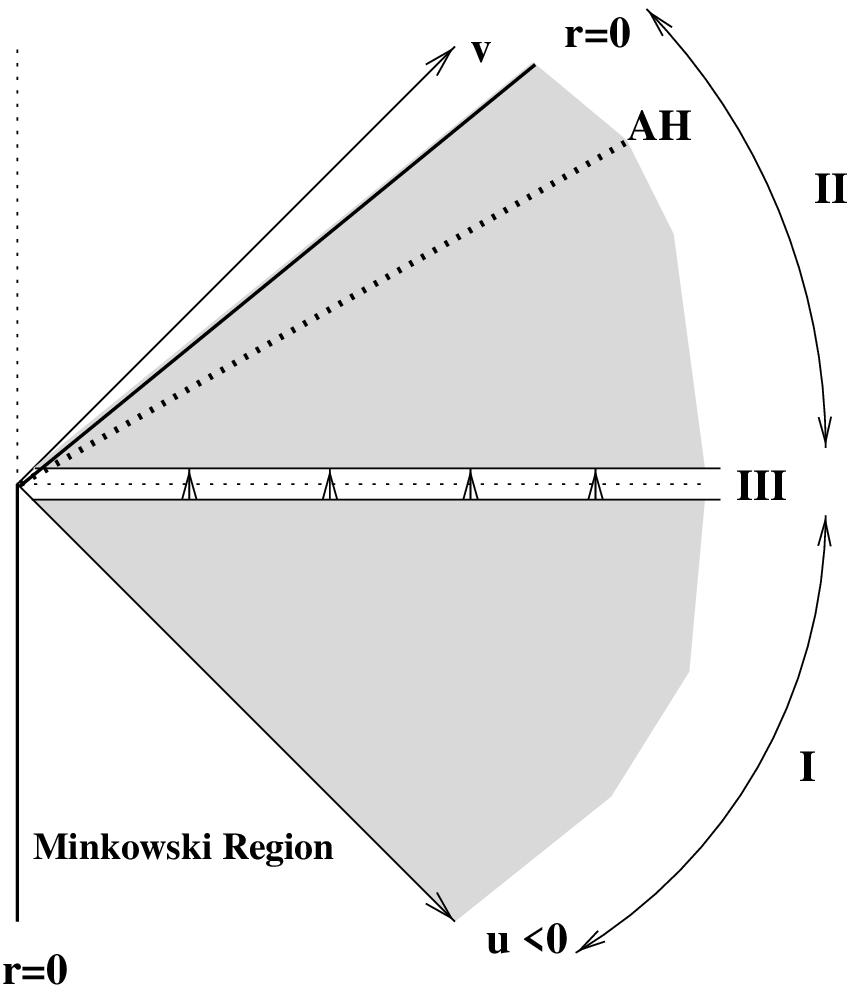}
}
%{\small\caption{Propagators}}
\vspace{.1in}
{\small Figure~2:~The spacetime diagram is obtained by joining the 
regions I and II. The presence of the region III is indicated by 
the strip with arrow, whose full structure does not appear in the diagram
 due to its Euclidean nature.}
\end{figure}  
 
At the initial boundary $(v=0,x=-\infty)$ we choose the 
integration 
constant of $\phi$ such that $\phi(v=0)=0$, and the 
stress-energy tensor 
components are 
\begin{equation} 
T_{uu}|_{v=0}=0, ~~T_{vv}|_{v=0} = 
\frac{1}{u^{2}(1-c_{0}^{2})},~~T_{uv}=0\,.
\label{eq15}  
\end{equation} 
At the final boundary $(u=0,x=+\infty)$ the components of 
stress-energy tensor are 
\begin{equation} 
T_{uu}|_{u=0}=\frac{1}{v^{2}(1-c_{0}^{2})},~~T_{vv}|_{u=0}=0,~~ 
T_{uv}=0\,,
\label{eq16}  
\end{equation} 
which implies that the spacetime is flat for 
$u>0$. However we will not consider the full region of $-\infty\leq 
x\leq\infty$, but limit to the region $-\infty\leq x\leq0$, and 
connect this to the Euclidean tunneling solution with a smooth 
boundary condition as shown in Fig. 2.
\vskip 0.4cm
(B) SOLUTION IN THE LORENTZIAN REGION II $(0<y\leq y_{II})$
\vskip 0.2cm 
In the region II where $y$ is restricted as 
$0<y\leq y_{II}$ the solutions to 
the equations (\ref{eq09}) and (\ref{eq10}) are 
\begin{eqnarray} 
&&y^{2}=1-\sqrt{1-c_{0}^{2}}\cosh2x\,,
\label{eq17}\\  
&&\phi=\frac{1}{2c_0}\ln\Biggl(\frac{\Gamma-e^{2x}}
{\Gamma^{-1}-e^{2x}}\Biggr) 
-\frac{1}{c_0}\ln \Gamma\,,
\label{eq18}  
\end{eqnarray} 
where $0<\Gamma=\sqrt{(1-c_{0})/(1+c_{0})}<1$ and 
$0<c_{0}<1$. 
The maximum of $y$ occurs at $x=0$ such that 
\begin{equation} 
y^{2}_{II}=1-\sqrt{1-c_{0}^{2}},~~x=0\,.
\label{eq19}  
\end{equation} 
There occurs black hole singularities when 
$y=0\,\,(r=y\sqrt{-uv}=0)$. Here, $\cosh2x_{0}=(z_{0}+
z^{-1}_{0})/2=1/\sqrt{1-c_{0}^{2}}$, 
and so 
\begin{equation} 
z_{0}=e^{2x_{0}}=\Gamma ~~or~~ \Gamma^{-1}\,.
\label{eq20}  
\end{equation} 
The past singularity occurs at $z_{0}=\Gamma$, while the future 
one at $z_{0}=\Gamma^{-1}$.
 
Notice that the geometry does not extend to the light 
cones $(u=0,\mbox{and}~ v=0)$ 
as the past light cone $(v=0,z=0)$ is located at the z-value 
smaller than the allowed limit $z_{0}=\Gamma$, and the future 
light cone $(u=0,z=\infty)$ has larger z-value than the 
allowed limit $z_{0}=\Gamma^{-1}$. 
There occurs apparent horizons where $(\bigtriangledown r)^{2}=0$ 
which implies $\dot{y}+y =0$. At the apparent horizon 
$y^{2}_{AH}={c_{0}^{2}/ 2}$, therefore 
\begin{equation} 
z_{H}=\sqrt{1-c_{0}^{2}}~~ \mbox{or}~~ \frac{1}{1-c_{0}^{2}}\,,
\label{eq21}  
\end{equation} 
where the first one is the past apparent horizon of the white 
hole, while the second one is the future apparent 
horizon of the black hole to be formed. 
\vskip 0.4cm
(C) SOLUTION IN THE EUCLIDEAN REGION $(y_{II}\leq y\leq y_{I})$ 
\vskip 0.2cm
In this region the geometry is no longer Lorentzian, but becomes 
Euclidean, which is classically forbidden. We consider this region 
as a tunneling mediated solution, and for this purpose we have to 
change the signature of the geometry and rederive the field 
equations. Fortunately we only need to substitute the scale 
variable $z$ as a complex number, namely, 
\begin{equation} 
z=e^{+2x} \longrightarrow z=e^{i2\theta}\,,
%~~~x~ 
%\mbox{and}~ \theta~ \mbox{are real} 
\label{eq22} 
\end{equation} 
where $\theta$ is the polar angle in the $u-v$ plane.
 Denoting $\dot{y}=dy/d\theta,~\dot{\phi}=d\phi/d\theta$ 
we obtain a new set of equations 
\begin{eqnarray} 
&&\dot{y}^{2}=-y^{2}+2 - \frac{c_{0}^{2}}{y^2}\,,
\label{eq23}\\   
&&\dot{\phi}= i \frac{c_{0}}{y^2}\,,
\label{eq24}  
\end{eqnarray} 
where we note that the tunneling potential $V_{\rm tun}(y)$ is an 
upside down flip of the Lorentzian potential $V(y)$ in Eq. (\ref{eq11}). 
 
The solution to the Eqs. (\ref{eq23})-(\ref{eq24}) are 
\begin{eqnarray} 
&&y^{2}=1+\sqrt{1-c_{0}^{2}}\cos2\theta,~~~0<c_{0}<1\,,
\label{eq25}\\   
&&\phi = \frac{i}{c_{0}}\tan^{-1}\Biggl(\frac{c_0\tan \theta}
{1+\sqrt{1-c_{0}^{2}}}\Biggr)+\phi_{0}\,.
\label{eq26}  
\end{eqnarray} 
We note that $y^2$ is periodic in $\theta$ as it should 
be because 
it represents a metric in a plane. At $\theta=0,~ y^2=1+
\sqrt{1-c_{0}^2}$ 
which is the same value as $y^{2}_{I}$ in the Lorentzian region 
I, so the Euclidean and the Lorentzian sectors are continuously 
connected. At $\theta=\pi/2,~y^{2}=1-\sqrt{1-c_{0}^{2}}$ 
and this is equal to the $y^{2}_{II}$ in the inner region II, here 
again continuous patching of geometry is possible.
 
\section{BLACK HOLE FORMATION BY TUNNELING} 
 
In the subcritical case collapsing scalar 
fields approach only to $y_{I}$ as $x$ approaches to zero 
starting from $x=-\infty~(v=0)$, and bounce back to the future 
light cone $(u=0)$. However, this collapsing scalar field may 
tunnel through the 
Euclidean region and appear again with the value $y_{II}$ at the 
matching point with the inner Lorentzian region. For this purpose 
we patch the three solutions to form a continuous complex 
spacetime as shown in the 
Fig. 2. 
 
We recall that in the Lorentzian region I and II, 
\begin{eqnarray} 
ds^{2}=-2dudv + r^{2}d\Omega^{2} 
=2dw^{2}-2w^{2}dx^{2}+r^{2}d\Omega^{2}\,,
\label{eq28}  
\end{eqnarray} 
where 
\begin{eqnarray} 
&&u=-we^{-x},~v=we^{x},~w>0\,,
\label{eq29} \\  
&&r=wy=y\sqrt{-uv}\,.
\label{eq30}  
\end{eqnarray} 
The solutions to be patched are 
\begin{eqnarray} 
y^{2}&=&1+\sqrt{1-c_{0}^{2}}\cosh2x,~-\infty<x\leq 0, {\rm\ \  (Region 
\ I)}\,, 
\label{eq31a}\\ 
y^{2}&=&1-\sqrt{1-c_{0}^{2}}~\cosh2x,~0\leq x<\infty, {\rm \ \ 
(Region\ 
II)}\,.
\label{eq31b}  
\end{eqnarray} 
Notice that at $x=0$ these solutions are not continuously matched. 
What we interpret on this mismatch is that, as we can see in the 
Fig. 2, as $x$ approaches zero the geometry come to the classical 
turning point and it tunnels through the potential barrier along 
the imaginary time while the real time remains at $x=0$. 
In the imaginary 
time, which is resulted by the Wick rotation of the time $x$, the 
signature of spacetime becomes Euclidean. Here we use 
\begin{equation} 
ds^{2}=2dw^{2}+2w^{2}d\theta^{2}+r^{2}d\Omega^{2}\,,
\label{eq32} 
\end{equation} 
where $0\leq \theta\leq 2\pi$, and $x$ in Eq. (\ref{eq28}) is substituted 
by $i\theta$.
The solution in this region is 
\begin{equation} 
y^{2}=1+\sqrt{1-c_{0}^{2}}\,\cos2\theta\,.
\label{eq33}  
\end{equation} 
Note that $y^{2}$ in the region I and this region 
continuously joins 
%this region 
at $x=0,~\theta=0$, their value being 
$y^{2}_{I}=1+\sqrt{1-c_{0}^{2}}$. Now the geometry evolves 
through $\theta$ until $\theta$ approaches $\pi/2$, where $y^{2}$ 
approaches to the value $y^{2}_{II}=1+\sqrt{1-c_{0}^{2}}\,$, 
where it patches to 
the region II continuously at $x=0$. 
 
As shown in Fig.~2, in the spacetime region prior to $v=0$ the 
geometry is flat Minkowski, and it evolves along $x$ from $x=-\infty$ 
to $x=0$ as the scalar fields collapse, then parts of the fields 
bounce back. But some may tunnel through the potential barrier 
evolving through the Euclidean geometry from $\theta=0$ to 
$\theta=\pi/2$, and the fields reappear at the inner 
region where the spacetime geometry forms an apparent horizon at 
$x_{AH}=-{1\over 4}\ln(1-c_{0}^{2})$, and finally collapse 
to form a black hole at $x_{BH}={1\over 4}\ln\Bigl((1+c_{0})/(1-c_{0})\Bigr)$ 
leaving a singularity $(r=0)$. 
 
In order to evaluate the probability to form a black hole by 
quantum tunneling through the Euclidean barrier we follow usual 
methods of instanton action calculation. We recall that 
the action 
is 
\begin{eqnarray} 
S_{sph} & = & {1\over 4}\int d^{2} x \sqrt{-\gamma} r^{2} 
[R(\gamma)+\frac{2}{r^{2}} \{(\bigtriangledown r)^{2}+1\} 
- 2 (\bigtriangledown \phi)^{2}] \nonumber \\ 
        & = & \int \omega d\omega d\theta ((\bigtriangledown 
        r)^{2} +1 - r^{2}(\bigtriangledown \phi)^{2})
\label{eq34}  
\end{eqnarray} 
where we use the metric, and the coordinates 
\begin{equation} 
\gamma_{ab} = \left( \begin{array}{cc} 
2 & 0 \\ 
0 & -2 \omega^{2} 
\end{array}\right), ~~u=-\omega e^{x},~~v=\omega 
e^{-x}\,.
\label{eq35}  
\end{equation} 
Since $r=\omega y$,  we can evaluate $(\bigtriangledown r)^{2}$ 
and $(\bigtriangledown \phi)^{2}$ 
as 
\begin{eqnarray} 
&&(\bigtriangledown r)^{2} = \gamma ^{ab} \partial_{a} r 
\partial_{b} r = \frac{1}{2}(y+\omega \frac{\partial}{\partial 
\omega}y)^{2} - \frac{1}{2} \dot{y}^{2}\,,
\label{eq36}\\   
&&(\bigtriangledown \phi)^{2} = \frac{1}{2} 
(\partial_{\omega}\phi)^{2}-\frac{1}{2\omega^{2}}\dot\phi^{2}\,, 
\label{eq37} 
\end{eqnarray} 
where $\dot{y}=\frac{\partial 
y}{\partial x},~\dot{\phi}=\frac{\partial \phi}{ \partial x}$. The  
action reads explicitly
\begin{equation} 
S_{sph}=\frac{1}{2}\int dq dx 
\left[-\dot{y}^{2}+y^{2}\dot{\phi}^{2}+2- y^2
+(2 q\frac{\partial}{\partial q}y)^{2}-y^{2}
(2q\frac{\partial}{\partial q}\phi)^{2}\right]\,.
\label{eq38}  
\end{equation}
where  $q\equiv {\omega^2\over 2}$ 
and we have dropped an irrelevant total derivative term. 
 
Now we turn to the particle picture tunneling through the 
potential barrier as shown in the Fig. 1. Defining the momentum 
densities as 
\begin{equation} 
\pi_{y}=\frac{\partial {\cal L}}{\partial \dot{y}} =  -
\dot{y},~~\pi_{\phi}=\frac{\partial {\cal L}}{\partial 
\dot{\phi}}=y^{2} \dot{\phi}\,,
\label{eq39}  
\end{equation} 
we obtain the effective superspace Hamiltonian density
\begin{eqnarray} 
{\cal H} & = & \pi_{y}\dot{y}+\pi_{\phi}\dot{\phi}-{\cal L} \nonumber \\ 
  & = & -\frac{\pi_{y}^{2}}{2}+\frac{\pi_{\phi}^{2}}{2 
  y^{2}}-\frac{1}{2}\left[2-y^2+(2q \frac{\partial}{\partial 
  q} y)^{2}-y^{2}(2q\frac{\partial}{\partial 
  q}\phi)^{2}\right]\,.
\label{eq40}  
\end{eqnarray}

In order to get the equivalent particle picture, one may 
consistently set
${\partial\hat\phi\over \partial q}|\psi\rangle =
{\partial \hat{y}\over \partial q}|\psi\rangle =0$ by demanding 
$|\psi\rangle$ to be a function only of the 
zero-mode variables, $\tilde{y}\equiv \int dq \, y(q) /\int dq $ and
$\tilde{\phi}\equiv \int dq \phi(q) /\int dq $. Then diagonalizing 
$\pi_\phi$ by $|\psi \rangle =|\Phi \rangle  e^{i c_0 \int dq \phi}$
with ${\delta\over \delta\phi} |\Phi \rangle =0$, the problem 
is further reduced to  
\begin{eqnarray} 
&& H|\Phi\rangle = 0\nonumber\\  
&&  H=-{1\over 2 K}{\partial^2\over  \partial \tilde{y}^2}+{K\over 2}\left[
2-\tilde{y}^2-{c^2_{0}\over {\tilde{y}^2}}\right]\,,
\label{eq42}  
\end{eqnarray} 
where $K\equiv \int\! dq$.
The effective action of the equivalent particle is then 
\begin{eqnarray} 
S_{eff}=\frac{K}{2}\int dx  
\left[\dot{\tilde{y}}^{2}-2+\tilde{y}^{2}+
\frac{c^2_{0}}{\tilde{y}^2}\right]
%+\frac{1}{4}\int 
%d\omega d\theta 
%\left[\omega^{2}\frac{\partial}{\partial\omega}y^2+
%(\omega^{2}\frac{\partial}{\partial\omega}y)^{2}\right] \nonumber \\ 
%&=&\frac{1}{4}\int d\omega d\theta \omega 
%\left[\dot{y}^{2}+2-\frac{c_{0}^{2}}{y^{2}}-y^{2}\right]\,,
\label{eq43}  
\end{eqnarray} 
with the Hamiltonian constraint ${H}\simeq 0$.
%, in the last line, we used the solution 
%$\partial y / \partial \omega 
%=0$, after taking out the total divergence term 
%${\partial (\omega^{2}y^{2}) \over 4 \partial 
%\omega}$. 
Notice that the Euclidean version of this action  gives the equation 
of motion of the particle given by Eq. (\ref{eq23}). 
The tunneling probability can be evaluated within the standard WKB 
approximation scheme.
Inserting the 
solution $y^{2}=1+\sqrt{1-c_{0}^{2}}\,\cos 2\theta$ 
with $x=i\theta$, and integrating 
from $\theta=0$ to $\theta=\pi/2$ we get 
\begin{eqnarray} 
S_{eff}&=& {K} 
\int^{\frac{\pi}{2}}_{0} 
d\theta \left[\frac{(1-c_{0})^{2}\sin^{2} 2\theta}
{1+\sqrt{1-c_{0}^{2}}\cos2\theta}\right] 
\nonumber 
\\ 
&=&  \int d\left ({\omega^2\over 2}\right) \frac{\pi}{2}(1-c_{0})\,,
\label{eq44}  
\end{eqnarray} 
where we used $K=\int d\left({\omega^2\over 2}\right)$ in the last line.
It is not possible to use this effective action as it diverges 
unless we introduce a cut-off in the $\omega-$variable.
As usual in the self-similar solutions this kind of divergences is 
unavoidable, so we introduce an arbitrary cut-off even though this 
is not allowed in the strict sense of self-similar 
solution. This divergence is an unnatural artifact of 
self-similarity as the black hole mass also diverges. With 
such a cut-off $\omega_{c}$, we get the probability of 
tunneling, 
therefore of black hole production as 
\begin{eqnarray} 
P=  e^{-2 S_{eff}} 
=  e^{-{\omega_{c}^{2}\over 2}\pi(1-c_{0})}\,.
\label{eq45}  
\end{eqnarray} 
The probability of the bouncing without black hole formation 
is just $1-P$. 
We note that in the limit of critical case $(c_{0}\rightarrow 1)$ 
we get the expected value $P=1$. 

Here the consideration of relative probability to the 
bouncing case does not remove the cut-off dependence contrary 
to the case of, for example, the free energy in 
the Schwarzschild-Anti-de-Sitter relative to pure Anti-de-Sitter 
space\cite{hawking}.
In order to avoid this 
divergent action naturally we may have to solve the 
equations with scalar 
flux of finite duration only, but this would require numerical 
studies.  One 
may interpret the solution as a near-critical solution with a 
finite mass and a finite probability of tunneling 
in the spacetime region of a scale  much below 
the relevant large correlation length. Then the cut-off scale 
will be 
naturally controlled by the correlation length. 
 
Brady\cite{brady} defined a local mass function $m(u,v)$ by 
\begin{equation} 
1-\frac{m(u,v)}{r} = 2 g^{uv}r_{,u}r_{,v}
\label{eq46}  
\end{equation} 
which agrees with both the ADM and Bondi masses in the appropriate 
limits. We apply this formula to get the black hole mass formed 
after tunneling $(x\geq 0)$ as 
\begin{eqnarray} 
m(u,v)= \frac{c_{0}^{2}}{4} \frac{\sqrt{-uv}}{y} 
= \frac{c^2_{0}}{4} 
\left(\frac{-uv}{1-\sqrt{1-c_{0}^{2}}\cosh2x}\right)^{\frac{1}{2}} 
\,,
\label{eq47}  
\end{eqnarray} 
and, along the apparent horizon 
$(z_{H}=1/\sqrt{1-c_{0}^{2}},~y^{2}_{AH}=c_{0}^{2}/2)$, 
we get 
\begin{equation} 
m_{AH}=\frac{v}{2\sqrt{2}}c_{0}(1-c_{0}^{2})^{\frac{1}{4}}\,,
\label{eq48}  
\end{equation} 
which agrees with the Brady's result\cite{brady}. In order 
to compare with 
the Brady's calculation, one should note that his 
parameters$(\alpha , \beta)$ 
are related to $c_{0}$ as $\alpha = 
\beta={1\over 2} {\sqrt{1-c_{0}^{2}}}$ 
giving an exponent ${1\over 2}$. 

The discrepancy of the critical exponent from the Choptuik's ($\sim 0.37$)
may be understood as follows. In the near-critical interpretation with
a finite but large correlation length scale, the apparent horizon will 
continue to develop in the outside region of the correlation length scale,
and eventually match with the event horizon if one considers the influx of 
scalar field with a finite duration. The mass viewed in this outside
 region, then, is expected to change appreciably and may 
conform with the Choptuik's.

\section{DISCUSSION} 
Frolov\cite{frolov} extended the work of Roberts\cite{robert} and 
Brady\cite{brady} to the 
problems in $n$-dimensions, where the action is 
\begin{equation} 
S=\frac{1}{16\pi}\int d^{n}x \sqrt{-g}
 [R-2(\bigtriangledown 
\phi)^{2}]\,,
\label{eq49}  
\end{equation} 
which reduces to the spherically symmetric action as 
\begin{equation} 
S_{sph} \propto \int d^{2}x \sqrt{-\gamma} r^{n-2} 
[R(\gamma)+(n-2)(n-3)r^{-2}((\bigtriangledown 
r)^{2}+1)-2(\bigtriangledown \phi)^{2}]\,.
\label{eq50}  
\end{equation} 
With the self-similarity ansatz we introduce 
\begin{equation} 
d\gamma^{2}=-2e^{2\sigma(z)}dudv,~r=-u\rho(z),~\phi=\phi(z)\,,
\label{eq51}  
\end{equation} 
and we get the effective particle motion as 
\begin{eqnarray} 
&&\dot{y}^{2}=y^{2}-2+c_{1}y^{-2(n-3)}\,,
\label{eq52} \\ 
&&\dot{\phi}= c_{0}y^{-(n-2)}\,,
\label{eq53}  
\end{eqnarray} 
where 
\begin{equation} 
c_{1}=\frac{2c_{0}^{2}}{(n-2)(n-3)} > 0\,.
\label{eq54}  
\end{equation} 
Here as in the previous cases we use the definitions,
%\begin{equation} 
$x=\frac{1}{2}\ln z$, and $\rho=\sqrt{z}y(x)$.
%\,,
%\label{eq55}  
%\end{equation} 
%and dots denote the derivative with respect to $x$. 
The solution can be formally presented as 
\begin{equation} 
x=\pm\int\frac{dy}{\sqrt{y^{2}-2+c_{1}y^{-2(n-3)}}} + c_{2}\,,
\label{eq56}  
\end{equation} 
which can be performed explicitly in terms of elliptic functions 
only in the dimension $n=5$, and $n=6$. 
 
In the Euclidean sector, the solution becomes 
\begin{equation} 
\theta = \int \frac{dy}{\sqrt{2-y^{2}-\frac{c_{1}}{y^{2(n-3)}}}} 
+c_{2}\,,
\label{eq57}  
\end{equation} 
where $0\leq \theta \leq \pi$. The solution $y=y(\theta)$ is an 
instanton like solution. 
We may join the above solutions continuously to form a tunneling 
geometry. Even though the solution need not be periodic, it is an 
interesting observation that the period satisfies the condition, 
\begin{equation} 
\frac{2\pi}{\sqrt{2(n-2)}} \leq T_{n} < \pi\ \ \ (n\ge 5)\,.
\label{eq58}  
\end{equation} 
Notice that the period is $\pi$ for the four dimensional spacetime, 
but 
there are no periodic solutions between $5\leq n \leq 9$, and yet 
there exists periodic solutions for $n \geq 10$. We do not know 
whether this is somehow connected with supergravity models of 
dimension ten or eleven. 
 
As a passing remark we mention that the 
effective potential (\ref{eq23}), 
$V(y)=y^{2}+c_{0}^{2} / y^{2} -2$, 
 is a potential of a three dimensional simple harmonic 
oscillator with an angular momentum $c_{0}$. It also 
reminds us the Calogero model\cite{olhanetsky} 
whose Hamiltonian is 
$H=p^{2} + \lambda^{2}/q^{2}$. Our potential may 
be viewed as simple harmonic oscillator 
modified by Calogero potential such that solubility 
might be 
preserved. 
 
Finally we should remind readers that our quantum black hole 
formation process is different from those black hole pair 
creations via instanton where gravitational collapse is not the 
driving mechanism\cite{bousso}. Our geometry is more kin to the 
Hartle-Hawking's 
no boundary creation of the Universe\cite{hartle}, or Vilenkin's 
tunneling 
proposal\cite{vilenkin}: our solutions
with reversed time describes nothing but the creation of the 
self-similar universe by tunneling, although its relevance
to cosmology needs to be clarified further.

\begin{acknowledgements} 
This work was supported in part by 
%the Basic Science
% Research 
%Institute
BSRI 
Program under  
BSRI 98-015-D00061, 
%the Korea Research Foundation 
KRF
under 98-015-D00054,
KOSEF Interdisciplinary Research Grant 98-07-02-07-01-5,
and  
%the Korea Science and Engineering 
%Foundation 
KOSEP through SNU-CTP and
under 98-07-02-02-01-3. 
One of us (Soh) would like to 
thank Korea Research Center for Theoretical Physics and Chemistry. 
\end{acknowledgements} 
%Work supported in part by KOSEF Interdisciplinary Research Grant and 
%SRC-Program, Ministry of Education Grant BSRI 98-2418 and 
%98-2425, SNU Faculty 
%Research Grant and Korea Foundation for Advanced Studies Faculty Fellowship.
 

\begin{references} 

\bibitem{choptuik}
M. W. Choptuik, Phys. Rev. Lett. {\bf 70}, 9 (1993).
\bibitem{christodoulou}
D. Christodoulou, Commun. Math. Phys. {\bf 105}, 337 (1986); 
{\bf 106}, 587 (1986); 
{\bf 109}, 591 (1987).   
%%%%%%%%%%%%%%%%%%%%%%%%%%%%%%%%%%%%%%%%%%%%%%%%%%%?????????????
\bibitem{goldwirth}
A. Goldwirth and T.Piran, Phys. Rev. D {\bf 36}, 3575 (1987); 
R. G\'{o}omez, R. A. Isaacson and J. Winnicour, 
J. Comput. Phys. {\bf 98}, 11 (1992).
\bibitem{robert}
M. D. Roberts, Gen. Rel. Grav. {\bf 21}, 907 (1989).
\bibitem{brady}
P. R. Brady, Class. Quantum Grav. {\bf 11} 1255 (1994).
\bibitem{frolov} 
A. V. Frolov, 
Class. Quant. Grav. {\bf 16} 407 (1999), gr-qc 9806112.
\bibitem{hawking} 
S.W. Hawking and  Don N. Page, Commun. Math. Phys. 
{\bf 87} 577 (1983). 
\bibitem{olhanetsky}
M. A. Olhanetsky and A. M. Perelomov, Phys. Rep. {\bf 71} 313 (1981).
\bibitem{bousso} 
R. Bousso and S. W. Hawking, Phys. Rev. D {\bf 52} 5659 (1995).
\bibitem{hartle} 
J. B. Hartle and S. W. Hawking, Phys. Rev. D {\bf 28} 2960 (1983), gr-qc 
9608009.
\bibitem{vilenkin}
A. Vilenkin, {\sl The quantum cosmology debate}, gr-qc 9812027. 
\end{references}
\end{document}